\documentclass[10pt,conference]{IEEEtran}

\usepackage[utf8]{inputenc}
\usepackage[T1]{fontenc}
\usepackage{cite}
\usepackage{amsmath}
\usepackage{amssymb}
\usepackage{url}
\usepackage{hyperref}
\usepackage{booktabs}
\usepackage{graphicx}
\usepackage{enumitem}
\setlist{topsep=2pt,itemsep=1pt,parsep=0pt}
\usepackage{xspace}
\usepackage{flushend}

\newcommand{\cfort}{CYBERFORT\xspace}
\newcommand{\cbridge}{CYBER-BRIDGE\xspace}

\newif\ifampmeasured
\ampmeasuredfalse 

\newif\ifsycmeasured
\sycmeasuredfalse 

\begin{document}

\title{The Amplifier Effect: Human-Factor Risks of\\
       AI-Suggested Correlation and Auto-Propagation\\
       in Multi-Framework GRC Self-Assessment}

\author{%
\IEEEauthorblockN{%
N.~Kekatos\textsuperscript{1},
M.~Ioannou\textsuperscript{2},
M.~Korgiala-Karyda\textsuperscript{3},
A.~Lekidis\textsuperscript{3},
T.~Nianios\textsuperscript{1}%
}
\IEEEauthorblockA{%
\footnotesize
\parbox{0.95\textwidth}{\centering
\textsuperscript{1}Clone Systems, Cyprus -- \texttt{\{nkekatos,\,tnianios\}@clone-systems.com}\quad
\textsuperscript{2}Bolton Technologies, Cyprus -- \texttt{michael@bts.com.cy}\quad
\textsuperscript{3}University of Thessaly, Greece -- \texttt{alekidis@uth.gr,\,korgialamarina@gmail.com}%
}}}

\maketitle
{\renewcommand{\thefootnote}{}\footnotetext{\copyright~2026 IEEE. Personal use of this material is permitted. Permission from IEEE must be obtained for all other uses, in any current or future media, including reprinting/republishing this material for advertising or promotional purposes, creating new collective works, for resale or redistribution to servers or lists, or reuse of any copyrighted component of this work in other works. This work has been accepted for publication at the 2026 IEEE International Conference on Cyber Security and Resilience (IEEE CSR 2026).}}

\begin{abstract}
Multi-framework Governance, Risk and Compliance (GRC)
platforms increasingly automate the link between an
organisation's self-assessment answer and the
compliance obligations that answer is said to
satisfy. Cross-framework control mapping, AI-suggested
question correlation, and automatic propagation of
answers and evidence across correlated questions all
serve the legitimate efficiency goal of reducing
duplicate work for small and medium-sized
enterprises (SMEs) under the EU Cyber Resilience Act
(CRA), NIS2 and GDPR. The same mechanisms,
however, \emph{amplify} the consequences of any
human-factor bias in a single answer: one
optimistically-graded control, one rubber-stamped
attestation, or one AI-drafted answer can be
silently replicated as evidence of compliance with
many obligations across multiple frameworks. We
call this the \emph{amplifier effect}: a
platform-design property (coarse-grained
attestation and un-gated propagation) rather than
a failing of individual users.
Using two EU-funded SME-facing GRC platforms,
\cfort and \cbridge, as examples, we (i) describe
the amplification mechanism in concrete data-model
terms, (ii) propose a six-dimension scoring
framework for evaluating any GRC tool's exposure to
the effect, (iii) instantiate the framework on a
thirteen-tool comparison covering enterprise IRM,
mid-market platforms, compliance-automation tools,
and the two EU SME projects, and (iv) outline a
measurement protocol that a consortium with access
to production self-assessment data can run. The thirteen-tool comparison is a structured design
assessment, not an empirical measurement of user
behaviour. The EU SME platforms score lowest on the
amplifier dimensions because their burden-reduction
design deliberately trades sign-off granularity for
throughput; we report this as a design trade-off, not
a verdict on the platforms. Our contribution is the
framing and the measurement protocol.
\end{abstract}


\IEEEpeerreviewmaketitle

\section{Introduction}
\label{sec:intro}
The EU regulatory stack for cyber resilience is
shifting from periodic external audit toward
\emph{continuous self-assessment} as the default
evidence of compliance. The Cyber Resilience Act
(CRA, Regulation~(EU)~2024/2847)~\cite{eu_cra_2024}
obliges manufacturers of products with digital
elements to perform and update a cybersecurity risk
assessment over the product lifecycle; NIS2
(Directive~(EU)~2022/2555)~\cite{nis2directive}
requires essential and important entities to adopt
risk-management measures proportional to exposure;
DORA places equivalent recurring duties on the
financial sector~\cite{dora2022}; the EU AI
Act~\cite{aiact} adds AI-system risk management to
the same regulatory scope. Each of these regimes
ultimately depends on \emph{the truthfulness and the
consistency of an organisation's own answers}.

Two structural responses to this shift are now
visible in the GRC tool market. First, modern
GRC/IRM platforms~\cite{grc_market_2025} converge on
a \emph{multi-framework} data model: a single
control or evidence object is mapped against several
frameworks simultaneously, so an SME does not redo
the same work for CRA, ISO/IEC~27001, NIS2, and
GDPR~\cite{scf}. Second, AI assistance is now embedded inside
the assessment workflow itself: LLM-drafted policies,
AI-suggested mappings between questions, AI-generated
readiness narratives, and retrieval-augmented
advisory~\cite{llm_compliance}.

Both responses are legitimate gains in
\emph{efficiency} (though, as we argue in
\S\ref{sec:disc}, not necessarily in usability, which
is the broader property). They
also, individually and especially in combination,
create a measurement problem that the human-factor
literature has not yet named. We call it the
\textbf{amplifier effect}: the property that, in a
modern multi-framework GRC platform, a single human
assessment action can drive many compliance
assertions, across many frameworks, with no
proportional increase in human review.

Framed precisely, the amplifier effect is less a
story about unreliable humans than about
\emph{design choices that create the conditions
for human error}: an attestation surface too coarse
to localise a mistake, a propagation step with no
confirmation gate, and an AI layer whose suggestions
are hard to tell apart from verified fact. It is a
domain-specific instance of the \emph{error-cascade}
problem studied in automated form-filling and
decision support~\cite{goddard2012}; what makes the
GRC setting acute is
that the cascade terminates in regulatory assertions,
so a general design failure becomes a compliance
failure.

\subsection{Concrete amplification mechanism}
Two EU-funded platforms make the mechanism easy to
inspect. \cfort~\cite{cyberfort2026} is an
open-source CRA-first GRC platform for SMEs deployed
under the Digital Europe Programme.
\cbridge~\cite{cyberbridge2026} is a cross-border
cybersecurity, compliance and forensics platform
under the same programme. Both share a
\emph{question-correlation} layer that, when a user
answers a question, finds all transitively
correlated questions inside the same scope and
\emph{propagates the user's answer and supporting
evidence to every correlated question
automatically}.\footnote{Implemented as
\texttt{synchronize\_correlated\_answers} and
\texttt{copy\_evidence\_files} in
\texttt{question\_correlation\_service.py}, verified at
commit \texttt{3ec8b9d} of the \cbridge code base; the
same service is present in \cfort.} The
correlation graph is itself partly authored by
humans and partly suggested by an AI service that
proposes new correlations between questions. The
resulting causal chain is:
\begin{quote}
\textit{(user action)} answers one question
$\;\rightarrow\;$
\textit{(platform)} finds the transitive correlation
group $\;\rightarrow\;$
\textit{(platform)} writes the same answer onto $N$
correlated questions $\;\rightarrow\;$
\textit{(platform)} copies the evidence object onto
$N$ correlated answers $\;\rightarrow\;$
\textit{(output)} $N$ compliance claims across one
or several frameworks.
\end{quote}
Each link is defensible in isolation. The
\emph{amplified} object (one human action, $N$
compliance claims) is what the human-factor
literature on self-assessment reliability has not
yet had to confront.

\subsection{Why the human-factor lens matters}
Self-assessment reliability is already known to be
limited by social-desirability bias~\cite{paulhus1991},
Dunning--Kruger~\cite{kruger1999}, the
expert--non-expert gap~\cite{ion2015}, and optimism
bias under organisational
incentives~\cite{adams-sasse,beautement2008}. The
HFaCS workshop topic list explicitly names
\emph{user compliance}, \emph{adoption barriers},
\emph{organisational culture} and
\emph{socio-technical root-cause analysis} as open
questions. The amplifier effect cuts across all four:
it converts an individual answer-level bias into an
organisation-level compliance posture with no
intermediate review gate, and it does so under a UI
that presents the propagation as a productivity
feature rather than as a chain of inferences. The AI
suggestion layer adds a second amplification: when
the correlation graph itself is partly AI-authored,
automation bias~\cite{skitka2000} affects which
propagations ever get scrutinised by a human.

\subsection{Contributions}
We do not run a controlled user study; we make a
design-level risk explicit and show how it would be
measured. Concretely: (1)~we name and define the
\emph{amplifier effect} as a property of
multi-framework GRC platforms with cross-question
answer propagation, and identify the four code-level
mechanisms that produce it (transitive correlation,
answer synchronisation, evidence propagation, and
AI-suggested new correlations); (2)~we propose a
\textbf{six-dimension scoring framework}
(\S\ref{sec:framework}) for evaluating a tool's
exposure: attestation surface, AI-assistance depth,
sign-off granularity, favourability-bias mitigation,
training-transfer measurement, and
self-report--telemetry corroboration; (3)~we
instantiate it on a thirteen-tool comparison
(\S\ref{sec:matrix}) spanning enterprise IRM,
mid-market, compliance-automation and quantification
tools plus \cfort and
\cbridge~\cite{grc_market_2025}; and (4)~we outline a
\textbf{measurement protocol} (\S\ref{sec:protocol}) a
consortium can run on production data, including five
empirical questions.

The remainder of the paper reviews self-assessment
reliability and the cross-framework GRC data model
(\S\ref{sec:background}), describes the amplifier
mechanism in code-level detail
(\S\ref{sec:mechanism}), introduces the six-dimension
scoring framework (\S\ref{sec:framework}) and the
thirteen-tool comparison (\S\ref{sec:matrix}), sketches
the measurement protocol (\S\ref{sec:protocol}), and
discusses design recommendations and threats to
validity (\S\ref{sec:disc}).

\section{Background}
\label{sec:background}

\subsection{Self-Assessment Reliability}
Self-assessment of cybersecurity posture inherits
the same biases that decades of psychometric work
have documented in survey research.
\textbf{Social-desirability bias}~\cite{paulhus1991}
pushes respondents toward answers that reflect well
on themselves or their organisation, particularly
when the assessment is tied to a regulatory
obligation. The \textbf{Dunning--Kruger
effect}~\cite{kruger1999} predicts that the
respondents least equipped to evaluate a control
are also the least aware of their own gaps;
Ion~et~al.~\cite{ion2015} show this concretely in
security, where non-experts misjudge protective
behaviours relative to expert benchmarks.
\textbf{Optimism bias under organisational
incentives}~\cite{adams-sasse,beautement2008}
predicts further inflation when a favourable answer
reduces individual workload; Beautement~et~al.\
formalise this as the \emph{compliance budget}, a
finite quantity of effort employees will spend on
security tasks before declining further work.
\textbf{Automation bias}~\cite{skitka2000} predicts
that respondents over-rely on a system-supplied
answer even when independent evidence is
available, an effect that is sharpened when the
system speaks with the authority of an AI
assistant~\cite{buccinca2021ai}.

In a workflow where each question is reviewed
independently, these biases bound the reliability
of one answer at a time. In a workflow where one
answer drives many, the same biases bound the
reliability of many answers at once.

\subsection{The Multi-Framework GRC Data Model}
Cross-framework mapping is a core capability of
modern integrated risk-management (IRM)
suites~\cite{grc_market_2025}. ServiceNow IRM, Archer
and MetricStream maintain harmonised control
catalogues mapped to ISO/IEC~27001, NIS2, NIST CSF,
SOC~2, GDPR, DORA and the CRA. Compliance-automation
tools (Drata, Vanta, Secureframe, Hyperproof) provide
similar mappings tuned to SOC~2 and ISO~27001
attestations. EU-funded SME projects, of which
\cfort and \cbridge are two~\cite{cyberfort2026,cyberbridge2026},
ship CRA-first mappings that overlap with
ISO/IEC~27001, NIS2 and GDPR. The reuse pattern is
the same across the market: a single control or
evidence object satisfies many obligations.

Continuous Control Monitoring (CCM) via
integrations closes part of
this gap by replacing periodic self-attestation with
telemetry-driven evidence. CCM coverage is, however,
uneven: it works well for technical controls
(identity, endpoint, cloud) and poorly for
organisational controls (training, awareness,
governance), which is where the workshop's
\emph{adoption} and \emph{organisational culture}
topics live. The amplifier effect emerges precisely
where CCM does not reach.

\section{Related Work}
\label{sec:related}

\textbf{Self-report in security.} The HAIS-Q
instrument~\cite{haisq} and SeBIS~\cite{sebis}
measure information-security awareness and intended
behaviour through Likert self-report; both have
documented social-desirability ceilings. Pfleeger
and Caputo~\cite{pfleeger2012} review behavioural
science approaches to security risk. Bada,
Sasse and Nurse~\cite{bada2019} document why
awareness campaigns relying on self-attestation
fail to change behaviour. The present paper extends
this line by examining how the modern GRC stack
\emph{multiplies} the consequences of any one such
answer.

\textbf{AI-assisted decision-making.}
Bu\c{c}inca, Malaya and Gajos~\cite{buccinca2021ai}
show that cognitive forcing functions can reduce
overreliance on AI in decision-making, an effect
relevant to the AI-suggestion layer in
\S\ref{sec:mechanism}. Skitka, Mosier and
Burdick~\cite{skitka2000} establish the
accountability--automation-bias trade-off that
underpins our recommendation~R2. We use the
term \emph{favourability bias} (sometimes called
\emph{sycophancy} in the LLM literature) to denote
the tendency of an AI assistant to produce
overly favourable answers when prompted to advocate
for the user.

\textbf{Compliance critique.} Herley's
analysis~\cite{herley2009} frames non-compliance
as a rational rejection of advice; the compliance
budget of Beautement~et~al.~\cite{beautement2008}
formalises the finite effort employees will spend
on security work. The amplifier effect extends this
line: a favourable answer with amplified
consequences yields a cheaper local outcome than
careful per-question scrutiny.

\textbf{GRC tool evaluation.} Gartner's public
market listing of integrated risk-management and
GRC tooling~\cite{grc_market_2025} provides a
vendor-neutral segmentation of GRC/IRM tooling and
underpins our thirteen-tool selection
(\S\ref{sec:matrix}). Existing comparisons evaluate
GRC platforms on functional breadth and on TCO; we
add a human-factor dimension to that comparison.

\section{The Amplifier Mechanism}
\label{sec:mechanism}

\begin{figure}[t]
\centering
\includegraphics[width=\columnwidth]{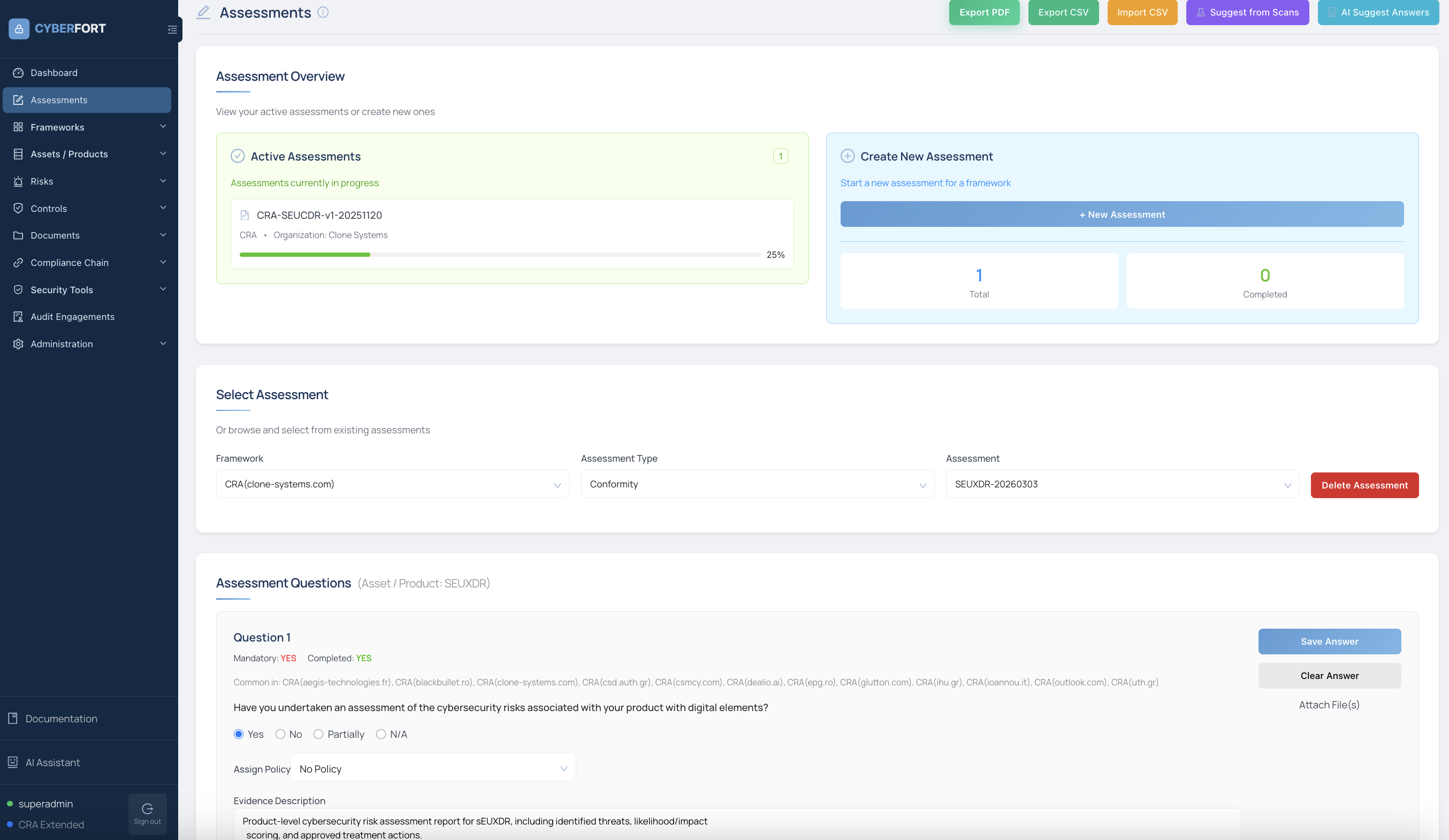}
\caption{Assessment view of \cfort. The
\textsc{Common in} line below the question lists
the correlated assessments to which the user's
answer will be propagated automatically by
\texttt{synchronize\_correlated\_answers}; the
\textsc{AI Suggest Answers} action at the top right
invokes the AI-suggestion layer. The two
mechanisms produce the amplifier effect described
in this paper.}
\label{fig:amplifier}
\end{figure}

\subsection{Walk-Through}
Both platforms treat a single \emph{answer} as the
unit of attestation: a value (\texttt{Yes},
\texttt{No}, \texttt{Partial}, \texttt{Not
Applicable}) plus an attached evidence document, tied
to one question in one assessment. When the user
answers one question and saves, which is the only
human action in the chain, a stored graph of question
correlations drives four mechanisms that the user
never reviews individually (Fig.~\ref{fig:amplifier}).

\paragraph{Transitive correlation}
The platform gathers every question transitively
correlated with the answered one in the same scope:
if $q_a$ is correlated with $q_b$ and $q_b$ with
$q_c$, all three form one group $G(q)$.

\paragraph{Answer synchronisation}
The same value is written onto every question in
$G(q)$ across the organisation's open assessments.
The user saw one save; the system records up to
$|G(q)|$ answers.

\paragraph{Evidence propagation}
The evidence document from the original answer is
copied onto every synchronised answer, so one document
now substantiates all $|G(q)|$ claims with no further
review.

\paragraph{AI-suggested correlations}
An ``AI Suggest'' action asks the assistant to propose
new correlation pairs. Accepted suggestions enlarge
$G(q)$, so the reach of the next single answer grows
with no human review of the propagation itself.

\subsection{Numerical Sensitivity}
Let $|F|$ be the number of frameworks active for the
organisation and let $\bar k$ be the mean
cross-framework correlation degree of a question
(empirically, $\bar k \in [2,5]$ for CRA-anchored
question banks that also cover ISO/IEC~27001 and
NIS2). The amplification factor of a single
user answer is then
\begin{equation}
    A(q) = |G(q)| \;\approx\; 1 + \bar k \cdot (|F|-1),
    \label{eq:amplification}
\end{equation}
which for $|F|=4$ and $\bar k = 3$ gives $A \approx
10$. One human attestation supplies the evidentiary
basis for ten compliance claims.

\subsection{A Worked Example}
Consider an SME running \cfort against CRA,
ISO/IEC~27001, and NIS2 simultaneously. Suppose the
correlation graph contains the following edges,
generated partly by human authors and partly by
the AI suggestion service:
\begin{itemize}[leftmargin=*]
    \item $q_{\text{CRA-13a}}$: ``Does the
    cybersecurity risk assessment cover known
    vulnerabilities?'' $\leftrightarrow$
    $q_{\text{ISO-8.8}}$: ``Is technical
    vulnerability management implemented?''
    \item $q_{\text{ISO-8.8}}$ $\leftrightarrow$
    $q_{\text{NIS2-21.2e}}$: ``Are vulnerability
    handling and disclosure measures in place?''
    \item $q_{\text{NIS2-21.2e}}$ $\leftrightarrow$
    $q_{\text{CRA-Annex-I-2.1}}$: ``Are products
    placed on the market without known exploitable
    vulnerabilities?''
\end{itemize}
The transitive closure groups all four questions.
A user answering $q_{\text{CRA-13a}}$ as
\texttt{Yes} and attaching an OSV-Scanner report as
evidence triggers
\texttt{synchronize\_correlated\_answers} to write
\texttt{Yes} onto the other three answers, and
\texttt{copy\_evidence\_files} to attach the
OSV-Scanner report to each. The platform emits a
single audit-trail entry per database row, but the
human-factor surface is one decision. If the user
mis-classified the scope of the OSV-Scanner run
(e.g., it covered only one of three product
components), all four claims inherit the error
silently.

\subsection{The \cbridge Variant}

\begin{figure}[t]
\centering
\includegraphics[width=\columnwidth]{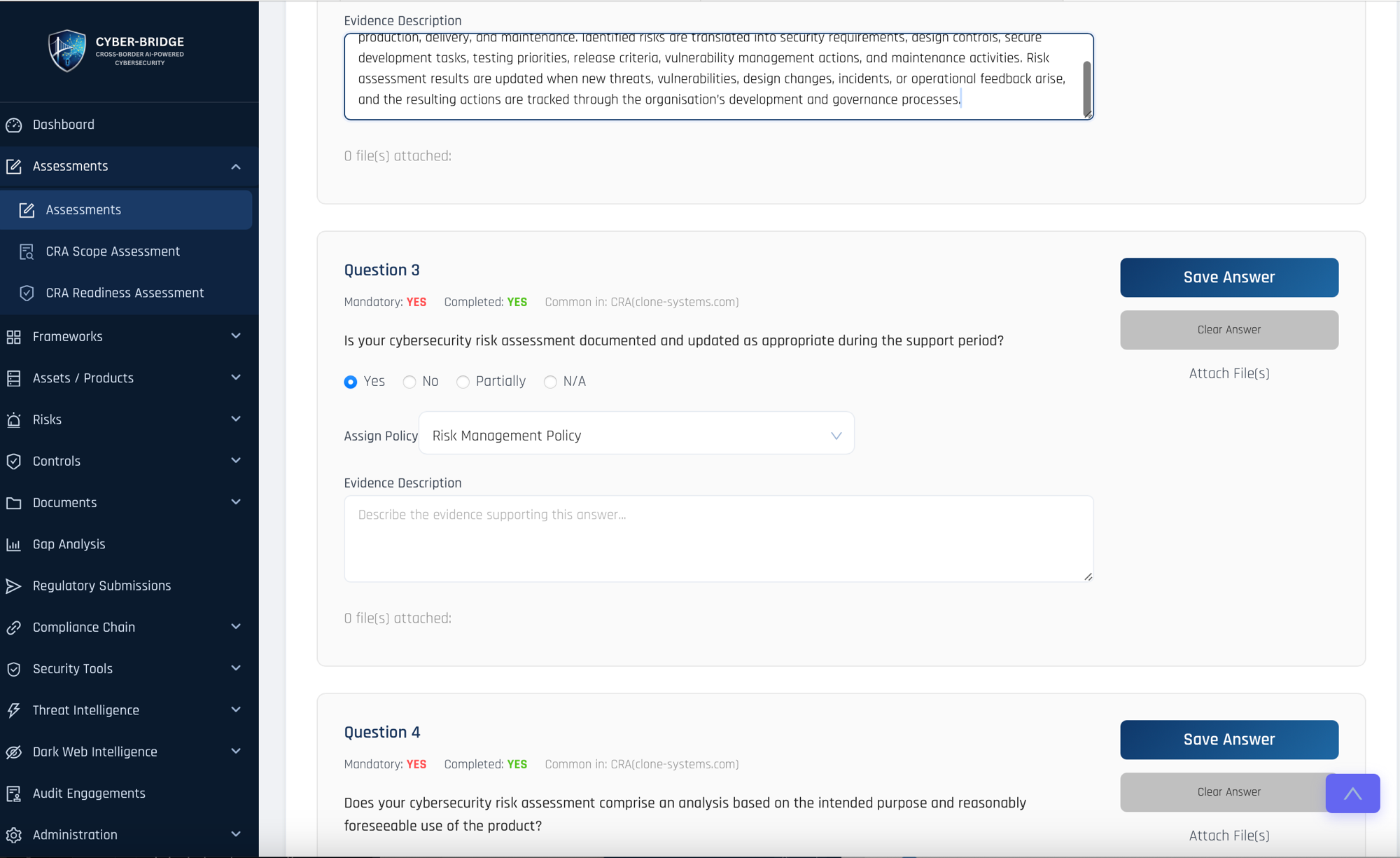}
\caption{Assessment view of the \cbridge platform.
The same propagation pattern observed in \cfort is
visible here: the \textsc{Common in} line below
each question lists the tenant assessment(s) to
which the user's answer is automatically copied,
and the assigned policy plus evidence description
are propagated alongside the value. The sidebar
shows the broader scope of the platform
(CRA scope, readiness, gap analysis, regulatory
submissions, threat intelligence) over which the
same amplifier mechanism operates.}
\label{fig:cbridge}
\end{figure}

\cbridge runs the same correlation and propagation
mechanism as \cfort, and adds a retrieval-augmented AI
assistant on top of it
(Fig.~\ref{fig:cbridge}).\footnote{A \texttt{pgvector}
embeddings store over the question bank, a local
\texttt{phi-4} model, and a regulatory-change search
layer.} The assistant is a \emph{second} amplification
path: it retrieves the $K$ questions most similar to
the one being answered, drafts answers for all of
them, and shows them in a single review pane; when the
user accepts, $K$ answers update at once. \cbridge is
thus exposed to the amplifier effect through both the
explicit correlation graph and the retrieval
assistant, yet the user sees the same thing: one
attestation, many claims.

\section{Six-Dimension Scoring Framework}
\label{sec:framework}

We score any GRC platform on six dimensions of
exposure to the amplifier effect. Each dimension
admits a four-level rubric: \textsc{None},
\textsc{Light}, \textsc{Moderate}, \textsc{Strong},
where higher is better. The per-tool aggregate in
Table~\ref{tab:matrix} is the rounded mean across the
six dimensions, scoring \textsc{None}${=}0$,
\textsc{Light}${=}1$, \textsc{Moderate}${=}2$,
\textsc{Strong}${=}3$.

\begin{enumerate}[leftmargin=*]
    \item \textbf{Attestation surface (AS):} fraction
    of controls corroborated by Continuous Control
    Monitoring (CCM) rather than self-attested;
    \textsc{Strong} $=$ CCM covers technical and most
    organisational controls, \textsc{None} $=$ all
    self-attested.
    \item \textbf{AI-assistance depth (AI):}
    reverse-scored; \textsc{Strong} $=$ AI only drafts
    text the human must rewrite, \textsc{None} $=$ AI
    suggests, propagates and persists with no review
    gate.
    \item \textbf{Sign-off granularity (SG):}
    \textsc{Strong} $=$ each propagated answer needs
    its own sign-off, \textsc{None} $=$ one signature
    covers a whole framework instance.
    \item \textbf{Favourability-bias mitigation (SM):}
    whether the user can steer the AI toward
    favourable answers (\emph{sycophancy});
    \textsc{Strong} $=$ a fixed, audited system prompt
    the user cannot override, \textsc{None} $=$ the
    user can freely instruct the AI.
    \item \textbf{Training-transfer measurement (TT):}
    \textsc{Strong} $=$ post-training behavioural
    metrics feed back into the assessment,
    \textsc{None} $=$ only completion is recorded.
    \item \textbf{Self-report--telemetry corroboration
    (RT):} \textsc{Strong} $=$ contradictions between
    answers and telemetry are surfaced, \textsc{None}
    $=$ no corroboration.
\end{enumerate}

\section{Thirteen-Tool Comparison}
\label{sec:matrix}

We instantiate the framework on thirteen
representative tools chosen to span the
size-and-maturity buckets named in the GRC market
analysis~\cite{grc_market_2025}: three enterprise IRM
suites, three mid-market platforms, four
compliance-automation tools, one specialised
quantification suite, and the two
EU-funded SME projects \cfort and \cbridge. Scoring
is drawn from public product pages and from the
market analysis itself, and from direct code
inspection for \cfort and \cbridge. The
tool comparison is a structured design assessment,
not an empirical measurement of user behaviour.
Scores should be read as indicative design-risk
ratings, not as audited vendor evaluations; the
evidence base for the EU SME rows is the inspected
development versions of the platforms and may
change as the projects mature.

\begin{table*}[t]
\centering
\caption{Thirteen-tool exposure to the amplifier effect
across six human-factor dimensions. Higher is better.
AS: attestation surface; AI: AI-assistance
mitigation; SG: sign-off granularity; SM:
favourability-bias mitigation; TT: training-transfer
measurement; RT: self-report--telemetry
corroboration.}
\label{tab:matrix}
\renewcommand{\arraystretch}{1.15}
\footnotesize
\setlength{\tabcolsep}{5pt}
\begin{tabular}{@{}lllllllll@{}}
\toprule
\textbf{Bucket} & \textbf{Platform} &
\textbf{AS} & \textbf{AI} & \textbf{SG} &
\textbf{SM} & \textbf{TT} & \textbf{RT} &
\textbf{Aggregate} \\
\midrule
Enterprise IRM         & ServiceNow IRM   & Strong   & Moderate & Strong   & Light    & Light    & Strong   & Moderate \\
                       & Archer           & Moderate & Light    & Strong   & Light    & Light    & Moderate & Moderate \\
                       & MetricStream     & Strong   & Light    & Moderate & Light    & Light    & Moderate & Moderate \\
\midrule
Mid-market             & LogicGate        & Moderate & Light    & Moderate & Light    & None     & Moderate & Light--Mod \\
                       & StandardFusion   & Moderate & Light    & Moderate & Light    & None     & Moderate & Light--Mod \\
                       & Onspring         & Light    & Light    & Moderate & Light    & None     & Light    & Light \\
\midrule
Compliance automation  & Drata            & Strong   & Light    & Moderate & None     & None     & Strong   & Moderate \\
                       & Vanta            & Strong   & Light    & Moderate & None     & None     & Strong   & Moderate \\
                       & Hyperproof       & Moderate & Light    & Moderate & None     & None     & Moderate & Light--Mod \\
                       & ControlMap       & Strong   & Light    & Moderate & None     & None     & Strong   & Moderate \\
\midrule
Quantification         & RiskLens         & Light    & Moderate & Moderate & Moderate & None     & Light    & Light--Mod \\
\midrule
EU SME projects        & \cfort           & Light    & None     & Light    & None     & Light    & Light    & Light \\
                       & \cbridge         & Light    & None     & Light    & None     & Light    & Moderate & Light \\
\bottomrule
\end{tabular}
\end{table*}

Three patterns emerge from Table~\ref{tab:matrix}.
\textbf{First}, enterprise IRM suites cluster at the
top of the range because they couple deep CCM
integrations (high AS, high RT) with mature sign-off
workflows (SG), but they remain weak on
training-transfer measurement (TT) and on
favourability-bias mitigation (SM).
\textbf{Second}, compliance-automation tools
maximise CCM but provide little human-attestation
structure: the auto-generated SOC~2 / ISO~27001
evidence packs come with weak SG and no SM, so a
single user can drive a complete framework
attestation. \textbf{Third}, EU SME projects \cfort
and \cbridge score lowest. This is by design: their
answer-propagation exists precisely to cut SME
workload, and that is the feature the amplifier
dimensions penalise. We flag our own tools as the
most exposed because we can inspect them directly,
which makes them the clearest place to measure the
effect. Their scanners (SAST/DAST/SBOM) corroborate
technical controls but not organisational ones such
as training, so self-report there goes unchecked
(RT~=~\textsc{Light}/\textsc{Moderate}).

The table describes where the risk sits; it is not
a ranking of good and bad tools. A \textsc{Light}
score is the expected result for a tool built to cut
SME workload, so it is not a failing in itself. The
real concern is narrower: when weak per-answer
sign-off (SG) is combined with AI writing the
answers, many compliance claims can rest on very
little human attestation, an outcome the regulation
did not anticipate because it assumed a human
meaningfully stands behind each claim.

\section{Measurement Protocol}
\label{sec:protocol}

A consortium that operates a production GRC
deployment (such as \cfort's cohort of $43$
SME tenants, $109$ users and $60$ framework
instances~\cite{cyberfort2026}) can answer five
empirical questions directly from its database,
under an aggregate-only publication policy. Each
question yields evidence on a different facet of
the amplifier effect.

\paragraph{Q1: answer distribution}
For each \texttt{FrameworkQuestion}, report the
empirical distribution of \texttt{Answer.value}
across all users. A skew toward favourable values
($\Pr[\text{Yes}] \gg 0.5$) on questions for which
no objective telemetry exists is the headline
optimism-bias signal.

\paragraph{Q2: time-to-answer}
Compute the per-answer dwell time as
$\Delta t_i = \texttt{updated\_at}_i -
\texttt{created\_at}_i$. A short
$\Delta t$ with a favourable \texttt{value} on a
correlated question is the headline rubber-stamp
signal.

\paragraph{Q3: evidence reuse multiplicity}
For each \texttt{Evidence} object linked to more
than one \texttt{Answer}, report the count.
A heavy-tailed distribution indicates that a small
number of evidence objects substantiate a
disproportionate share of the platform's
compliance claims: the single-point-of-failure
signal.

\paragraph{Q4: AI-assisted vs.\ unassisted}
Where the AI-assistance invocation is logged per
answer, compare the answer distribution and the
$\Delta t$ distribution between the AI-assisted and
unassisted strata. A leftward shift of $\Delta t$
without a corresponding shift in the answer
distribution is consistent with automation
acceleration without bias change; a shift in both
is consistent with automation bias.

\paragraph{Q5: favourability-bias probe}
For ten representative CRA questions, prompt the
platform's LLM (Mistral~7B for \cfort, phi-4 for
\cbridge) twice: once with a neutral-consultant
system prompt, once with an
``advise-the-SME-favourably'' prompt. Report
divergence on the answer rubric. A non-zero
divergence quantifies the sycophancy attack
surface of the AI-assistance layer.

The five questions are mechanically reproducible
once instrumentation is in place. Q1--Q3 require
only existing tables; Q4 may require adding a
per-answer AI-invocation log; Q5 is a one-day
controlled probe.

\section{Discussion}
\label{sec:disc}

\subsection{A Usability-Heuristic Reading}
The amplifier effect can be restated as a set of
violated usability heuristics~\cite{nielsen1994}.
Silent propagation breaches \emph{visibility of
system status}: the user is never shown that one
answer became $N$ compliance claims; the absence of a
per-target confirmation breaches \emph{user control
and freedom} and \emph{error prevention}; and
surfacing AI-suggested correlations
indistinguishably from human-authored ones erodes the
user's model of what the system actually knows. Read
this way, the effect is a predictable consequence of
omitting the confirmation, undo, and status-disclosure
affordances that the human-factors literature treats
as baseline, which is why we describe the
propagation as an efficiency gain but not a usability
one.

\subsection{Design Recommendations}
Three recommendations follow from the mechanism and
the matrix. \textbf{(R1) Review gate on propagated
answers.} The answer-synchronisation step should
require an explicit per-target acknowledgement when
the propagated answer would change a previously
filled value, or when the target question is
mandatory and the source is not.
\textbf{(R2) AI-action audit log.} Every AI
invocation that touches an answer, correlation, or
evidence record should be persisted with the
originating user ID, the system prompt, and a hash of
the AI output, to support both Q4 and external audit.
\textbf{(R3) Transfer measurement on training.}
Cyber-range deliverables in both \cfort and \cbridge
log completion but not behavioural transfer; a
mandatory post-training behavioural probe
(e.g., a phishing simulation rate measured before
and after) would convert the TT column from
\textsc{Light} to \textsc{Moderate}.

\subsection{Threats to Validity}
The scoring in Table~\ref{tab:matrix} is derived
from public sources and from direct code inspection
of two platforms. We did not conduct hands-on
evaluation of the eleven commercial tools; their
scores reflect documented features and may be revised
by trial use or vendor input, and a vendor may
legitimately dispute a \textsc{Light} score. Only
\cfort and \cbridge were assessed by direct
inspection.
The amplification factor in
\eqref{eq:amplification} is a structural upper
bound; the realised amplification depends on the
density of the correlation graph in deployment.
The measurement protocol is described but not
executed; we treat its execution as the natural
next deliverable for the \cfort and \cbridge
consortia.

\subsection{Ethical and Deployment Considerations}
Analysing production self-assessment data carries
workforce-monitoring implications. We assume the
five-question protocol is run under
aggregate-only publication, that no individual or
tenant-identifiable rows leave the consortium, and
that the SME tenants are informed of the analysis
in their original data-sharing agreement. The
sycophancy probe (Q5) is conducted on synthetic
vignettes rather than on tenant answers, and
poses no additional disclosure risk.

\section{Conclusion}
\label{sec:conc}

In this paper, we have introduced and characterised
the \emph{amplifier effect} in multi-framework
governance, risk, and compliance platforms: the
property that a single human self-assessment
action, combined with cross-framework mapping and
automatic answer/evidence propagation, drives many
compliance assertions without a proportional
increase in human review.
We have described the mechanism in code-level
detail for two EU-funded production platforms,
proposed a six-dimension scoring framework for
evaluating any GRC tool's exposure, instantiated
the framework on thirteen representative tools, and
outlined a five-question measurement protocol that
a consortium with access to production
self-assessment data can execute.
The amplifier effect is a structural property of
the modern GRC stack and is tractable to measure
once named. Future work will execute the
measurement protocol on the \cfort first cohort,
report the empirical distributions of $A(q)$ and
$\Delta t$ across the deployment, and complement the
log analysis with \emph{user research} (think-aloud
sessions and a controlled test of whether making the
propagation scope visible changes attestation
behaviour) to close the end-user perspective that
this design-level analysis leaves open.

\section*{Acknowledgment}
This work has received funding from the European
Union's Digital Europe Programme under grant
agreements No~101249702 (CYBER-BRIDGE) and
No~101190281 (CYBERFORT).


\begin{thebibliography}{99}

\bibitem{eu_cra_2024}
European Parliament and Council of the European
Union, ``Regulation (EU) 2024/2847 on horizontal
cybersecurity requirements for products with
digital elements (Cyber Resilience Act),''
\emph{Official Journal of the European Union},
Nov.~2024.

\bibitem{nis2directive}
European Parliament and Council of the European
Union, ``Directive (EU) 2022/2555 on measures for
a high common level of cybersecurity across the
Union (NIS2),'' \emph{Official Journal},
L~333, pp.~80--152, Dec.~2022.

\bibitem{dora2022}
European Parliament and Council of the European
Union, ``Regulation (EU) 2022/2554 on digital
operational resilience for the financial sector
(DORA),'' \emph{Official Journal}, L~333,
Dec.~2022.

\bibitem{aiact}
European Parliament and Council of the European
Union, ``Regulation (EU) 2024/1689 laying down
harmonised rules on artificial intelligence (AI
Act),'' \emph{Official Journal}, Jul.~2024.

\bibitem{cyberfort2026}
N.~Kekatos \emph{et~al.}, ``CYBERFORT: A
compliance-chain platform operationalising the
Cyber Resilience Act for SMEs,'' under submission,
IEEE CSR, 2026. Project page:
\url{https://cyber-fort.eu/}.

\bibitem{cyberbridge2026}
N.~Kekatos \emph{et~al.}, ``The CYBER-BRIDGE
project: an AI-powered platform for cross-border
cybersecurity, compliance and forensics in the
EU,'' under submission, IEEE CSR, 2026. Project
page: \url{https://cyber-bridge.eu/}.

\bibitem{grc_market_2025}
Gartner, ``Integrated risk management solutions
reviews and ratings,'' Gartner Peer Insights, 2025.
[Online]. Available:
\url{https://www.gartner.com/reviews/market/integrated-risk-management}
(accessed Jul.~2026).

\bibitem{paulhus1991}
D.~L.~Paulhus, ``Measurement and control of
response bias,'' in \emph{Measures of Personality
and Social Psychological Attitudes}, J.~P.~Robinson,
P.~R.~Shaver, and L.~S.~Wrightsman, Eds. Academic
Press, 1991, pp.~17--59.

\bibitem{kruger1999}
J.~Kruger and D.~Dunning, ``Unskilled and unaware
of it: how difficulties in recognizing one's own
incompetence lead to inflated self-assessments,''
\emph{Journal of Personality and Social
Psychology}, vol.~77, no.~6, pp.~1121--1134,
Dec.~1999.

\bibitem{ion2015}
I.~Ion, R.~Reeder, and S.~Consolvo,
``\ldots no one can hack my mind: comparing expert
and non-expert security practices,'' in
\emph{Proc.\ SOUPS}, 2015, pp.~327--346.

\bibitem{adams-sasse}
A.~Adams and M.~A.~Sasse, ``Users are not the
enemy,'' \emph{Communications of the ACM},
vol.~42, no.~12, pp.~40--46, Dec.~1999.

\bibitem{beautement2008}
A.~Beautement, M.~A.~Sasse, and M.~Wonham, ``The
compliance budget: managing security behaviour in
organisations,'' in \emph{Proc.\ NSPW}, 2008,
pp.~47--58.

\bibitem{skitka2000}
L.~J.~Skitka, K.~Mosier, and M.~D.~Burdick,
``Accountability and automation bias,''
\emph{International Journal of Human-Computer
Studies}, vol.~52, no.~4, pp.~701--717, Apr.~2000.

\bibitem{haisq}
K.~Parsons \emph{et~al.}, ``The human aspects of
information security questionnaire (HAIS-Q): two
further validation studies,'' \emph{Computers \&
Security}, vol.~66, pp.~40--51, 2017.

\bibitem{sebis}
S.~Egelman and E.~Peer, ``Scaling the security
wall: developing a security behavior intentions
scale (SeBIS),'' in \emph{Proc.\ ACM CHI},
2015, pp.~2873--2882.

\bibitem{pfleeger2012}
S.~L.~Pfleeger and D.~D.~Caputo, ``Leveraging
behavioral science to mitigate cyber security
risk,'' \emph{Computers \& Security}, vol.~31,
no.~4, pp.~597--611, 2012.

\bibitem{bada2019}
M.~Bada, A.~M.~Sasse, and J.~R.~C.~Nurse, ``Cyber
security awareness campaigns: why do they fail to
change behaviour?'' arXiv:1901.02672, 2019.

\bibitem{herley2009}
C.~Herley, ``So long, and no thanks for the
externalities: the rational rejection of security
advice by users,'' in \emph{Proc.\ NSPW}, 2009,
pp.~133--144.

\bibitem{buccinca2021ai}
Z.~Bu\c{c}inca, M.~B.~Malaya, and K.~Z.~Gajos,
``To trust or to think: cognitive forcing
functions can reduce overreliance on AI in
AI-assisted decision-making,'' in \emph{Proc.\
ACM Human-Computer Interaction (CSCW)},
vol.~5, no.~CSCW1, art.~188, 2021.

\bibitem{nielsen1994}
J.~Nielsen, ``Enhancing the explanatory power of
usability heuristics,'' in \emph{Proc.\ ACM CHI},
1994, pp.~152--158.

\bibitem{scf}
Secure Controls Framework Council, ``Secure Controls
Framework (SCF): a cybersecurity and privacy
metaframework mapping 1{,}500+ controls to 200+ laws,
regulations and frameworks,'' 2025. [Online].
Available: \url{https://securecontrolsframework.com/}
(accessed Jul.~2026).

\bibitem{llm_compliance}
S.~Hu, P.~Wang, K.~I.~Chan, Y.~Yao, and Z.~Lu,
``\emph{I always felt that something was wrong}:
understanding compliance risks and mitigation
strategies when highly-skilled compliance knowledge
workers use large language models,'' arXiv:2411.04576,
2024.

\bibitem{goddard2012}
K.~Goddard, A.~Roudsari, and J.~C.~Wyatt, ``Automation
bias: a systematic review of frequency, effect
mediators, and mitigators,'' \emph{Journal of the
American Medical Informatics Association}, vol.~19,
no.~1, pp.~121--127, 2012.

\end{thebibliography}
\end{document}